\documentclass[sigconf]{acmart}

\AtBeginDocument{%
  }

\copyrightyear{2026}
\acmYear{2026}
\setcopyright{cc}
\setcctype{by}
\acmConference[RecSys '26]{20th ACM Conference on Recommender Systems}{September 27-October 02, 2026}{Minneapolis, MN, USA}
\acmBooktitle{20th ACM Conference on Recommender Systems (RecSys '26), September 27-October 02, 2026, Minneapolis, MN, USA}
\acmDOI{10.1145/3773078.3831871}
\acmISBN{979-8-4007-2284-4/2026/09}

\begin{document}

\title{PinDCO: Whole-Page Aware Dynamic Creative Optimization at Scale}

\author{Yu Hao}
\affiliation{%
  \institution{Pinterest Inc.}
  \city{San Francisco}
  \country{USA}}
\email{yhao@pinterest.com}

\author{Yuchun Li}
\affiliation{%
  \institution{Pinterest Inc.}
  \city{Seattle}
  \country{USA}}
\email{yuchunli@pinterest.com}

\author{Peimeng Sui}
\affiliation{%
  \institution{Pinterest Inc.}
  \city{San Francisco}
  \country{USA}}
\email{psui@pinterest.com}

\author{Meilin Liu}
\affiliation{%
  \institution{Pinterest Inc.}
  \city{San Francisco}
  \country{USA}}
\email{meilinliu@pinterest.com}

\author{Tianyuan Cui}
\affiliation{%
  \institution{Pinterest Inc.}
  \city{San Francisco}
  \country{USA}}
\email{tianyuancui@pinterest.com}

\author{Hao Li}
\affiliation{%
  \institution{Pinterest Inc.}
  \city{San Francisco}
  \country{USA}}
\email{haoli@pinterest.com}

\author{Zicong Zhou}
\affiliation{%
  \institution{Pinterest Inc.}
  \city{San Francisco}
  \country{USA}}
\email{zzhou@pinterest.com}

\author{Akanksha Baid}
\affiliation{%
  \institution{Pinterest Inc.}
  \city{San Francisco}
  \country{USA}}
\email{abaid@pinterest.com}

\renewcommand{\shortauthors}{Hao et al.}

\begin{abstract}
Recent advances in generative AI have substantially accelerated the creation of high-quality ad creatives, dramatically expanding the number of candidate variants per campaign. This shift increases the need for scalable dynamic creative optimization (DCO) systems that can match creatives to the most relevant audiences under stringent latency and cost constraints. We present PinDCO, a production DCO system for ad creative retrieval and selection on Pinterest, a billion-scale visual discovery platform.

PinDCO is built around a Creative Component Fusion Network (CCFN) that performs dynamic creative scoring by modeling each creative component (e.g., image, title, layout) with a dedicated tower, using component-specific hyperparameters to account for differing modeling complexity. The component representations are fused to predict a creative-level score conditioned on the ad-level prediction, and we improve training data quality via an exploration-exploitation strategy.

To account for Pinterest's waterfall grid layout, where a creative's rendered size affects nearby content and session-level engagement, we introduce a Pixel-aware Adjustment Module(PAM) that adjusts scores based on creative size to encourage efficient screen real-estate utilization and better whole-page outcomes. To support the large volume of creative candidates, we further employ a lightweight pre-selection model for early pruning, and optimize serving efficiency through caching and dynamic batching. Extensive offline analyses and online A/B experiments demonstrate the effectiveness of PinDCO, yielding a $+3.09\%$ lift in ad Click-Through Rate(CTR) with positive whole-page metrics. With the strong performance, we launched PinDCO in the Pinterest Ads platform.

\end{abstract}




\maketitle

\section{Introduction}
\begin{figure}[ht]
  \centering
  \includegraphics[width=0.2\textwidth]{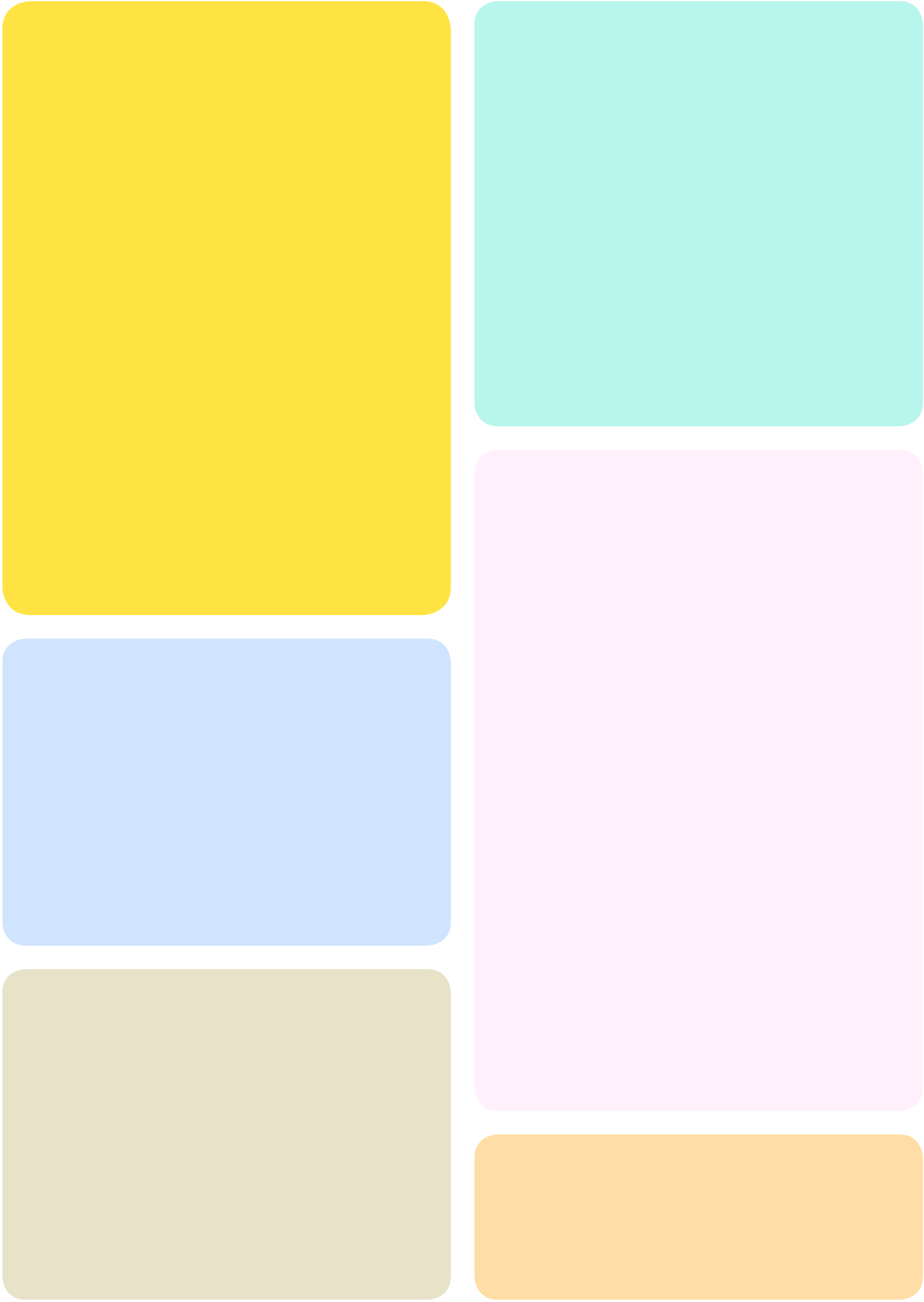}
  \caption{Illustration of the Pinterest waterfall grid layout. Each colored block denotes a grid cell that displays a creative. Grid cells are organized into vertical columns with a fixed width, while their heights vary.}
  \Description{Illustration of the Pinterest waterfall grid layout. Each colored block denotes a grid cell that displays a creative. Grid cells are organized into vertical columns with a fixed width, while their heights vary.}
  \label{fig:waterfall-grid}
\end{figure}

Pinterest is a visual discovery platform powered by personalized recommendations that help users find inspiration and ideas (e.g., recipes, home decor, fashion, and travel). Its interface adopts a waterfall grid layout, in which creatives are arranged in vertical columns. Creative quality is critical to user engagement, particularly in advertising settings \cite{maros2019image,zhao2019you,azimi2012visual,ge2018image}. Visually compelling creatives can signal a more trustworthy merchant with higher quality products. This work focuses on selecting ad creatives to improve advertising performance. 

For a given ad, there is rarely a single creative that performs best for all audiences. Dynamic Creative Optimization (DCO) \cite{lin2022joint,yang2024parallel} personalizes creative selection by considering not only visual quality, but also contextual signals such as user preferences and message relevance. Rather than adopting a one-creative-fits-all strategy, DCO allocates a diverse set of creatives to the audiences with which they resonate most, thereby improving overall campaign performance. For example, DCO can adapt creative selection as users progress through the shopping journey. On Pinterest, users with different intents benefit from different creative strategies: casual browsers often respond to visually native ads that blend with organic content and build brand awareness over time, whereas high-intent shoppers typically gain more value from creatives that present direct and relevant product information to support comparison and decision making. Accordingly, DCO selects the creative that best matches the current user and context under the ad's optimization objective.

Modern recommendation systems typically comprise multiple stages, including retrieval and ranking. To enable personalized creative selection within this pipeline, prior work either introduces a dedicated downstream stage for creative optimization \cite{lin2022joint} or parallelizes creative optimization with other stages \cite{yang2024parallel}. These architectural choices reflect trade-offs among model capacity, latency, and serving cost. In this work, we develop a scalable and versatile DCO system for the generative-AI era, tailored to a visually rich platform with unique practical constraints. 

The Pinterest waterfall grid layout introduces requirements that are not well served by conventional DCO formulations that optimize each ad slot independently. In the layout shown in Figure~\ref{fig:waterfall-grid}, creatives are arranged in vertical columns with a fixed width, but variable heights. Taller creatives occupy more screen real estate and push subsequent content downward within the same column, whereas shorter creatives leave additional room for content below. These cross-slot interactions motivate whole-page optimization that explicitly accounts for efficient use of screen pixels. Thus, beyond maximizing per-slot engagement, the system should regulate a creative's pixel footprint to preserve space for other ads and organic content.

The rapid progress of generative image models can cause creative repositories to grow exponentially, as automated generation is substantially cheaper and faster than manual design. This proliferation of candidates places significant pressure on online model serving. To limit additional end-to-end latency while supporting higher-capacity models, we parallelize DCO with the heavy ranking stage, following \cite{yang2024parallel}. Consequently, the winning ad is unknown at serving time, the system must score creatives for all ad candidates, resulting in extremely high QPS that can approach the billion scale and therefore requires careful efficiency optimizations to control cost.

The continual evolution of image generators also raises the bar for creative modeling. Modern creative generators incorporate feedback from real-world serving data and update over time to produce more engaging creatives \cite{yang2024new}, continuously introducing new assets for DCO to allocate. Such fresh content is particularly susceptible to cold-start issues \cite{zhao2019you}, as the system has limited historical signals. Exploration and exploitation \cite{wang2021hybrid} is a standard approach to rapidly gather informative feedback for new creatives while reducing exposure to underperforming variants. Exploration traffic provides less-biased observations, which is especially important for creative-level optimization, where subtle visual differences are difficult to infer from content features alone and can be better captured through user engagement signals. Moreover, high-quality exploration data helps models adapt to shifting distributions of creative assets and user preferences over time.

To address these challenges, we present PinDCO, a scalable and high-performance creative optimization system validated on production Pinterest traffic. We begin by constructing a high-quality training dataset using an exploration-exploitation strategy. Specifically, we apply $\epsilon$-greedy to ensure diverse traffic coverage across creatives and users. We then train a Creative Component Fusion Network (CCFN) to predict the incremental CTR (and/or other objectives) relative to the ad ranking prediction that's agnostic to creative variants. CCFN comprises multiple component-specific towers, each modeling a creative component (e.g., image, title, and layout). We assign different hyperparameters (such as dropout rates) across towers to reflect component complexity and to mitigate asynchronous convergence, where some towers underfit while others overfit. The model is jointly trained and calibrated with the base ad ranking model, only predicting the delta between ad-level score and creative-level score. It greatly lowers the difficulty of the learning task.

After scoring, a Pixel-aware Adjustment Module adjusts predictions as a function of creative size to encourage efficient utilization of screen pixels by penalizing overly tall creatives when meaningful. Under this mechanism, a creative must improve engagement without excessively degrading session-level outcomes for neighboring grid cells to win against other variants of the same ad. To maintain model capacity without increasing end-to-end latency, CCFN runs in parallel with ad ranking. Because scoring all creatives for all ad candidates is expensive at high volume, we introduce a lightweight pre-selection unit before CCFN. This unit runs locally in the ad service, consumes a small set of features directly from the creative serving index, avoiding expensive RPCs to the model server and feature store. It dynamically reduces the number of CCFN candidates to trade off infrastructure cost and performance opportunity. Finally, we improve serving efficiency via dynamic batching, sharded local caching on the model server, and cache memory tuning. PinDCO is deployed on the Pinterest advertising platform and delivers a significant $+3.09\%$ ad Click-Through Rate(CTR) lift with positive whole-page metrics under reasonable infrastructure cost.

We summarize our contributions as follows: 
\begin{itemize}
\item We propose PinDCO, a creative optimization system that incorporates a Pixel-aware Adjustment Module to account for whole-page impacts in addition to CTR.

\item We design a Creative Component Fusion Network for dynamic creative scoring, with component-specific towers and hyperparameter configurations. We further extend Exploration and Exploitation to an online personalized creative optimization setting, whereas previous work primarily utilize this strategy in offline non-personalized creative modeling.

\item We improve serving efficiency via lightweight early pruning, cache optimization, and dynamic batching to accommodate the large volume of creative candidates.

\item We deploy PinDCO on production Pinterest Ads traffic and observe a significant $+3.09\%$ CTR lift together with positive whole-page metrics. We also conduct extensive offline and online studies to characterize the contribution of individual techniques.
\end{itemize}

\begin{figure*}[ht]
  \centering
  \includegraphics[width=0.9\textwidth]{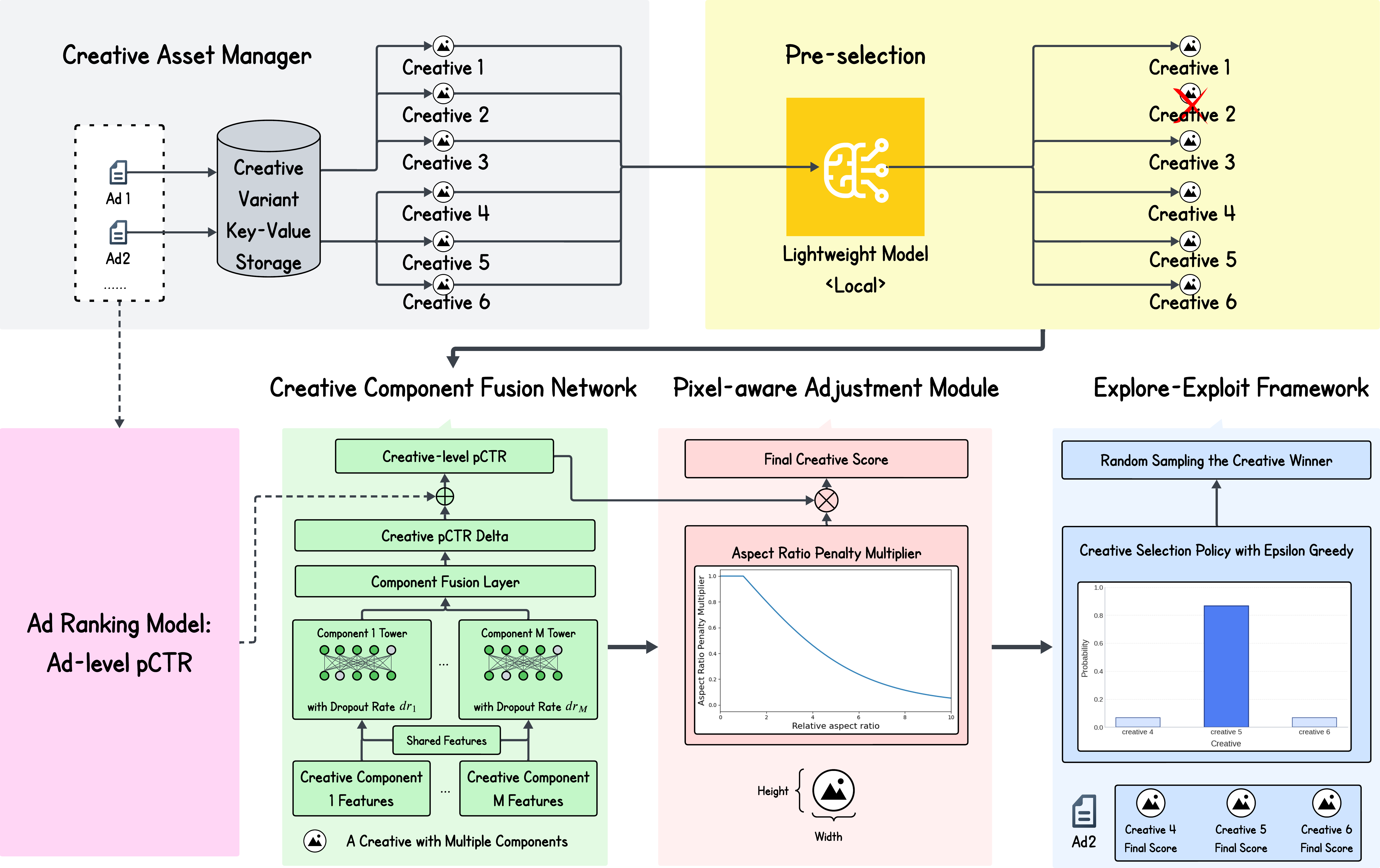}
  \caption{PinDCO system architecture. Given ads retrieved upstream, the ad service expands each ad into a candidate set of creatives by querying a key-value store, and prunes candidates using a local lightweight model to reduce downstream inference cost. The ad service then invokes the model server to run the Creative Component Fusion Network for creative scoring. A Pixel-aware Adjustment Module is applied to support whole-page optimization. Finally, we balance exploration and exploitation with a multi-armed bandit to collect higher-quality training data.}
  \Description{PinDCO system architecture. Given ads retrieved upstream, the ad service expands each ad into a candidate set of creatives by querying a key-value store, and prunes candidates using a local lightweight model to reduce downstream inference cost. The ad service then invokes the model server to run the Creative Component Fusion Network for creative scoring. A Pixel-aware Adjustment Module is applied to support whole-page optimization. Finally, we balance exploration and exploitation with a multi-armed bandit to collect higher-quality training data.}
  \label{fig:main-architecture}
\end{figure*}

\section{Related Work}
\subsection{Offline Creative Quality Scoring}
Creative quality is pivotal to advertising and e-commerce performance. Users are more likely to engage with creatives that feature aesthetically appealing visuals and coherent messaging across components (e.g., image and text). A common paradigm is to predict creative CTR(Click-Through Rate) from content features, ranging from individual components \cite{talebi2018nima, maros2019image, zhao2019you, you2023image} to interactions among components (layout, multi-image, graph representation) \cite{yang2019learning, fan2022automatic}. Beyond content signals, SoCraft\cite{huang2023socraft} debiased CTR prediction using campaign setup information, which is a key factor for ad CTR. While CTR is the most common target, this framework readily extends to other objectives, such as CVR(Conversion Rate). 

Differences among creatives are often subtle and difficult to capture using content features alone. Consequently, empirical user responses have proven critical for creative scoring. This setting typically requires balancing exploration, to collect unbiased and reliable feedback, with exploitation, to limit the cost of serving suboptimal arms. Verizon \cite{koren2020dynamic, aharon2019carousel} optimize carousel assets using a multi-armed bandit algorithm (Successive Elimination). VAM-HBM \cite{wang2021hybrid} combines empirical observations with visual priors when determining the creative winner. 

Creative quality scoring can steer generation for both people and models: it gives designers automated, peer-informed feedback to speed iteration \cite{kong2023neural, mishra2021tsi, verma2022recommendation, mishra2019guiding, zhou2020recommending, mishra2020learning}, and it acts as a critic to early-stop weak assets \cite{vempati2020enabling, youngmann2020automated, kanungo2022cobart}, generate RLHF-style labels \cite{yang2024new}, and guide efficient composition search to reduce exploration regret \cite{chen2021automated, chen2021efficient}.

\subsection{Online Creative Selection}
Usually no single creative is optimal for all audiences; personalized selection matches variants to segments using real-time user and context signals, and is typically modeled online.

Traditional multi-stage recommenders use a lightweight retrieval stage to fetch promising candidates coarsely, then a heavier ranker for fine-grained selection. A common strategy is to inject creative signals into the ranker: visual inputs are encoded \cite{azimi2012visual, ge2018image, liu2020category, xv2022visual, chen2022hybrid, yu2022boost, wen2024unified, li2020adversarial, tao2023event} alongside other ad-level features to improve CTR prediction, often requiring task-specific changes to the encoder and architecture.

However, this approach does not scale as creatives per ad grow. To avoid overwhelming the ranking model with an excessive number of candidates, CACS \cite{lin2022joint} adds a lightweight creative selection stage before ranking, improving CTR prediction but adding sequential serving latency. Peri-CR \cite{yang2024parallel} runs selection in parallel with ranking to reduce latency, yet still ranks every creative variant for all ad candidates.

\section{Methodology}

The PinDCO system comprises multiple units designed to address the aforementioned challenges. This section presents the high-level architecture and then describes each unit in detail.

\subsection{Architecture}
We build upon a conventional recommendation pipeline with an ad retrieval stage followed by an ad ranking stage. The ranking stage typically employs a large model with substantial latency and cost, making it impractical to evaluate a large number of creative variants. To avoid end-to-end latency degradation, we parallelize the creative optimization units with the ad ranking stage, leveraging the ranking latency to hide the creative optimization latency.

Along the creative optimization branch, we employ several units, as illustrated in Figure \ref{fig:main-architecture}. The creative asset manager retrieves creative metadata indices keyed by ad ID from a key-value store. The Pre-selection unit applies a lightweight model to reduce the candidate set for efficiency. The ad service then invokes the model inference server to run the Creative Component Fusion Network for creative score prediction, in parallel to the ad ranking model. The Pixel-aware Adjustment Module adjusts scores to account for whole-page optimization. Finally, the Multi-Armed Bandit unit overrides the predicted choice with some probability to collect randomized feedback.

\subsection{Creative Asset Manager and Pre-Selection}
\label{creative-asset-manager}
The creative asset manager takes as input the ad IDs returned by the retrieval stage, queries a creative-index key-value store, and expands each ad into a set of creative candidates. Given the high QPS and number of entries, we retain only a minimal set of creative metadata. With the reduced data footprint, we deploy a cache cluster to cache metadata aggressively and reduce load on the backend key-value store.

The pre-selection unit prunes low-opportunity creatives to reduce downstream cost, using a lightweight model (e.g, rules, multi-armed bandits or linear classifier) whose features come from the creative-index key-value store. It runs locally in the ad service, avoiding extra model server RPCs and infrastructure overhead.

Pre-selection aggressiveness is a key performance-cost lever: retaining more candidates often improves downstream performance but increases serving cost. In extreme cases where most creatives of an ad have low opportunity, the lightweight model can be configured to return a single candidate to bypass subsequent stages, substantially reducing cost.

\subsection{Creative Component Fusion Network}
A creative typically comprises multiple components, such as image, title, and layout. We propose a model that first learns component-specific representations and then fuses information across components to form a holistic representation of the creative.

\subsubsection{Model Architecture}
The model comprises three layers of building blocks: feature encoding, component-specific towers, and component fusion.

The feature encoding layer maps raw inputs to vector representations consumed by subsequent stages. It normalizes continuous features via clipping, whitening, or log-transformation depending on the input distribution. Categorical or ID features are mapped to embeddings using lookup tables. The resulting features are grouped by the creative component they describe, and features that are not tied to a specific component are placed in a "Shared" group. The main purpose is to handle different creative components that have heterogeneous feature spaces by routing each component’s features to a specialized tower, while still leveraging shared features across components. Each tower takes as input the corresponding component features together with the "Shared" feature group.

Finally, the output embeddings from all towers are combined in the component fusion layer, which aims to capture interactions across components. The fusion layer produces a creative delta score that is combined with the ad ranking model prediction. The sum of the two logits is supervised with the final user engagement label. This separation of responsibilities enables CCFN to focus on creative-level differences while relying on the ad ranking model to mitigate ad-level confounding factors. When the objective is to predict only the delta over the ad-level base engagement, CCFN can use a smaller feature set and a shallower network architecture than the ad ranking model.

\subsubsection{Features}
We have a few category of features:
\begin{itemize}
    \item Content Features: capture the visual and semantic information of the creative components. We use vector representations precomputed by Pinterest image and text encoders. In addition, we manually engineer features that incorporate domain knowledge to improve model performance. Examples include foreground-background color distributions, foreground product size, and image aspect ratio.
    \item User Features: support creative personalization. They include basic demographics and user understanding signals derived from prior activity, such as shopping intent and user embedding representations.
    \item Empirical Features: leverage historical observations of creative variants to predict their future performance. Empirical rates and counts are typically aggregated into buckets based on the slicing scheme of users and creative variants. Selecting an appropriate slicing scheme is critical. Smaller buckets aggregate over a more compact and fine-grained set of records, yielding more specialized estimates, but they are more susceptible to sparsity and noise. We therefore adopt a multi-scale aggregation strategy that combines fine-grained and coarse-level buckets.
    \item Context Features: describe the context in which the recommendation is made. The same user may prefer different creatives across contexts. For example, creative brightness can be adjusted by hour-of-day, and the optimal layout may differ between mobile and web.
\end{itemize}
\subsubsection{Creative-Component-Specific Hyperparameters}
\label{dropout-paragraph}
Each tower in the CCFN is responsible for modeling a single creative component. Components differ in both the number of variants and the complexity of modeling. For example, images typically contain higher-dimensional information than other components and are therefore more challenging to learn. If all towers are trained at the same pace, their convergence rates may differ. Some towers may begin to overfit while others remain underfitting.

To better align convergence across towers, we introduce creative-component-specific hyperparameters. For example, each tower uses a distinct dropout rate that reflects the characteristics of its corresponding component. These hyperparameters are tuned to stabilize the training loss and improve creative selection performance. 

\subsection{Pixel-aware Adjustment Module for Whole-Page Optimization}
A larger creative may improve the current ad's performance, but it can cannibalize user's attention for other ads or organic content. Beyond ad CTR, we also optimize for whole-page engagement across both ad and organic slots. The objective is to allocate screen real estate wisely to maximize effective pixel utilization across all content. Additional pixels should be assigned to an ad creative only when the marginal gain is sufficient to offset the cannibalization of other content.

We introduce a Pixel-aware Adjustment Module(PAM) for whole-page optimization. It imposes a penalty that increases with creative size, on the CCFN predicted score. This adjustment downweights longer creatives unless they provide sufficient performance gains. A creative must maximize incremental value while limiting its pixel footprint. We currently use aspect ratio to measure the creative size, assuming same grid width for all variants. But it can be extended to more complex scenarios where grid width also varies.

Consider an ad with original creative aspect ratio ${ar}_{orig}$ and a candidate creative with aspect ratio ${ar}_{cand}$. We define the relative aspect ratio as ${ar}_{rel} = \frac{{ar}_{cand}}{{ar}_{orig}}$. Let $S$ denote the final creative score, $O$ the CCFN output score, and $P({ar}_{rel})$ the aspect ratio penalty. With $k$ as a tunable penalty strength, we have
$$ S = O \cdot P({ar}_{rel})$$
$$ P({ar}_{rel}) = \operatorname{clip} (1 - \tanh(k({ar}_{rel}-1)),\, 0,\, 1) $$
Note that $P(\cdot)$ should be monotonically decreasing and take values in $[0, 1]$. We adopt a $\tanh$ form for simplicity, although alternative choices are possible. We leave the investigation of other penalty functions to future work. Figure \ref{fig:main-architecture} shows an example of the curve in the Pixel-aware Adjustment Module chart.

We use offline replay to guide the selection of the penalty strength $k$. Given logged model prediction scores and aspect ratios for all creative candidates of an ad, we compute the final creative scores and simulate winner selection. Summary statistics from the simulated winners are then used to tune $k$. For example, we can choose $k$ to achieve a target average aspect ratio.

\subsection{Exploration and Exploitation}
Rather than always selecting the winner greedily based on the Final Creative Score, we incorporate exploration to collect high-quality data with reduced selection bias. This is particularly important for creative optimization due to (1) the large volume of newly introduced creatives, (2) subtle differences that require empirical features to distinguish, and (3) continuously shifting user preferences. We used an $\epsilon$-greedy explore–exploit algorithm to balance performance needs against production constraints such as serving cost, latency, and observability overhead.

Let $\mathcal{C} = \{c_1, c_2, \dots, c_n\}$ denote the set of creatives, and let $S(c_i)$ be the Final Creative Score of creative $c_i$. Define the highest-scoring creative as $c^* = \arg\max_{c_i \in \mathcal{C}} S(c_i)$. Under the $\epsilon$-greedy algorithm, the policy of creative selection is

$$
\Pr(W = c_i) =
\begin{cases}
1-\varepsilon, & \text{if } c_i = c^*,\\[6pt]
\dfrac{\varepsilon}{n-1}, & \text{if } c_i \neq c^*.
\end{cases}
$$

The selected winning creative $W$ is served to the user, and engagement signals are collected for empirical feature aggregation and training data labeling.
 
\subsection{Serving Efficiency}
\label{Serving Efficiency}
We deploy the CCFN model on a dedicated serving cluster. Upon receiving a request containing metadata for candidate creatives from the ad service, the model server retrieves candidate features from an online feature store and runs inference. Given the high inference QPS, we apply the following optimizations:

\textbf{Caching Optimization:} We reduce load on the backend feature store by introducing a local cache on the model server. To maximize cache hit rate, we shard scoring requests by creative id so that each creative is consistently routed to the same machine. We also increase the cache memory threshold to retain as many entries as possible without degrading inference performance.

\textbf{Dynamic Batching:} Each ad request contains hundreds of ad candidates, and each candidate includes multiple creative variants. The ad service submits scoring requests to the model server with a list of candidates to score. Naively concatenating all candidates into a single request can substantially increase latency. The request must wait for all candidates to complete, which increases the likelihood that one slow candidate dominates end-to-end latency. It poorly parallelizes the request and response processing. To balance latency and throughput, we flatten the creative variants across all ad candidates and partition them into batches of tunable size. Batches are scored and processed in parallel to maximize machine utilization and minimize latency.

\section{Experiments}
In this section, we compare our approach with baseline methods, evaluate the contribution of individual system components, and summarize experimental findings. Our online experiments demonstrate significant metric improvements in A/B tests, and we subsequently launched PinDCO on the Pinterest Advertising Platform. Metrics are reported as relative changes with respect to the baseline.

\subsection{Offline Experiments}
We first evaluate our primary model, the Creative Component Fusion Network, using an offline dataset derived from Pinterest production traffic, and conduct ablation studies to quantify the contribution of each model component.

\textbf{Dataset and Metrics:} We build the dataset from several months of Pinterest production ad engagement logs. We focus on image and layout components due to their volume and impact, and plan to add more component types as creative generation expands. We use logged engagement as supervision, downsample negatives for class balance, and calibrate predicted scores accordingly. Following standard practice, we report ROC-AUC (Receiver Operating Characteristic area under the curve) and PR-AUC (Precision-Recall area under the curve) as evaluation metrics for our binary classifier.

\textbf{Baseline Models:} We compare our approach with several baselines to assess performance:
\begin{itemize}
\item No-CR: No creative ranking is applied. The creative variant is selected using a simple global rule. This reflects the production setting prior to the PinDCO launch.
\item Lightweight-Only: Section \ref{creative-asset-manager} describes a lightweight model that runs before the main CCFN model to reduce serving cost via early pruning. We evaluate its performance to ensure early pruning is guided by a sufficiently accurate model.
\item Peri-CR\cite{yang2024parallel}: A creative module is jointly trained with the ad ranking module and served in parallel. This parallel structure increases creative modeling capacity and supports stronger creative personalization. 
\end{itemize}

\subsubsection{Discussion on the offline results}

\begin{table}[h]
\centering
\begin{tabular}{lcc}
\hline
Model & PR-AUC & AUC-ROC \\
\hline
No-CR & - & - \\
Lightweight-Only &  +0.111\% & +0.027\% \\
Peri-CR & +0.163\% & +0.043\% \\
CCFN & \textbf{+0.171\%} & \textbf{+0.046\%} \\
\hline
\end{tabular}
\caption{Relative Percentage change of creative selection model PR-AUC and AUC-ROC compared against no-CR baseline.}
\Description{Relative Percentage change of creative selection model PR-AUC and AUC-ROC compared against no-CR baseline. The CCFN has the best performance compared to other models.}
\label{tab:offline_model_comparison}
\end{table}

Table \ref{tab:offline_model_comparison} summarizes the offline performance of different creative selection methods. Compared with the no-CR setting, all learned approaches improve creative selection quality, which confirms the value of incorporating creative-aware modeling into the selection process. The Lightweight-Only model yields a 0.111\% gain in PR-AUC and a 0.027\% gain in AUC-ROC, showing that even a lightweight selection model can capture useful signals for coarse creative discrimination. However, its relatively limited AUC-ROC improvement suggests that a small model is insufficient for fine-grained ranking among candidate creatives.

Peri-CR improves AUC-ROC by $0.043\%$, and PR-AUC by $0.163\%$ over no-CR, indicating that a dedicated creative modeling module can better capture personalized creative effects than a simple model. CCFN achieves the best overall performance, with a $0.171\%$ gain in PR-AUC and a $0.046\%$ gain in AUC-ROC. This result suggests that our model is more effective for distinguishing subtle differences among creative variants.

Overall, the offline results validate the effectiveness of the proposed CCFN architecture. They also support the broader system design in which a lightweight stage is used for efficient yet effective early pruning, while a higher-capacity model is reserved for fine-grained creative scoring. This combination provides a favorable trade-off between model quality and practical serving constraints.

\subsubsection{Ablation Study}

\begin{table}[h]
\centering
\begin{tabular}{lcc}
\hline
Model & PR-AUC Lift & AUC-ROC Lift \\
\hline
CCFN &  +0.171\% & +0.046\% \\
 w/o Exploration Data & +0.164\% & +0.042\% \\
 w/o Component-Specific Dropout & +0.163\% & +0.043\% \\
\hline
\end{tabular}
\caption{Ablation study: relative percentage change of PR-
AUC and AUC-ROC compared against no-CR, when a specific component is removed}
\Description{Ablation study: relative percentage change of PR-
AUC and AUC-ROC compared against no-CR, when a specific component is removed. Removing any of the components causes performance drop.}
\label{tab:ablation_study}
\end{table}

To understand the contribution of key design choices in CCFN, we conduct an ablation study and report the results in Table \ref{tab:ablation_study}. Specifically, we examine the effect of removing exploration data and the effect of removing component-specific dropout. When either of these components is removed, performance drops, indicating that both designs make consistent positive contributions to model quality.

Removing exploration data hurts performance because creative optimization depends on high-quality empirical signals (we keep the training data volume constant for a fair comparison). With many newly introduced and subtly different variants, content features alone are often insufficient; exploration traffic provides broader, less biased coverage that improves empirical aggregation and mitigates selection bias and cold-start effects.

Removing component-specific dropout also leads to worse performance. As described in Section \ref{dropout-paragraph}, different creative components exhibit different levels of modeling difficulty and may converge at different rates during training. Applying separate dropout rates to different towers helps balance these differences and stabilizes optimization. The ablation result suggests that this mechanism improves representation learning for component-specific towers and leads to better final creative scoring after fusion.

\subsection{Online Experiments}
To evaluate the end-to-end PinDCO system and its platform impact across ads and organic content, we deploy PinDCO in the Pinterest ads serving funnel. We use the following two key platform-level metrics for measurement:
\begin{itemize}
    \item Ad CTR: Click-through rate of all ads on the platform. It measures ad engagement and serves as a primary metric for ad models.
    \item Successful Session: Number of user sessions with success-defining actions, for example, click, save, or social interactions. It includes both ad and organic actions and captures overall whole-page-level impact.
\end{itemize}

\begin{table}[h]
\centering
\begin{tabular}{lcc}
\hline
Model & Ad CTR & Successful Session \\
\hline
No-CR & - & - \\
Lightweight-Only & +1.49\% & -0.24\% \\
Peri-CR & +1.70\% & -0.12\% \\
PinDCO & \textbf{+3.09\%} & \textbf{+0.04\%} \\
\hline
\end{tabular}
\caption{Relative percentage change of creative selection online performance compared against no-CR baseline.}
\Description{Relative percentage change of creative selection online performance compared against no-CR baseline. PinDCO outperforms all the baseline approaches in terms of CTR, while maintaining a slightly positive whole page metric.}
\label{tab:online_model_comparison}
\end{table}

Table \ref{tab:online_model_comparison} reports the online performance of different creative selection approaches relative to the no-CR baseline. All learned methods improve ad CTR, which confirms that creative selection can generate measurable gains in live traffic. The Lightweight-Only variant yields a +1.49\% lift in Ad CTR, indicating that a lightweight pre-selection model can realize part of the creative opportunity at relatively low serving cost. Peri-CR further improves Ad CTR to +1.70\%, while PinDCO achieves the largest gain at +3.09\%, substantially outperforming both baselines. These results are consistent with the offline findings and suggest that higher-capacity creative modeling yields additional value beyond lightweight filtering alone.

The two platform metrics together also highlight the importance of whole-page-aware optimization. Although both Lightweight-Only and Peri-CR improve Ad CTR, they lead to negative changes in Successful Session, at -0.24\% and -0.12\%, respectively. In contrast, PinDCO achieves a positive +0.04\% lift in Successful Session while also delivering the strongest Ad CTR improvement. This result suggests that improving local ad engagement alone is insufficient, and that creative selection should account for broader whole-page-level effects on user experience.

We attribute this improvement to the full system design of PinDCO. In addition to a stronger creative scoring model, PinDCO explicitly incorporates a Pixel-aware Adjustment Module to regulate pixel footprint under Pinterest’s waterfall layout, and uses exploration data to improve training quality for creative-level modeling. Together, these components enable the system to improve ad engagement while better preserving the quality of the surrounding page. Overall, the online experiment validates that PinDCO can deliver significant ad gains together with positive whole-page metrics under production constraints.
\subsubsection{Serving Efficiency Improvement}

\begin{table}[h]
\centering
\begin{tabular}{lcc}
\hline
System & Latency P99 & Latency P90 \\
\hline
PinDCO & - & - \\
w/o Dynamic Batching & +87\% & +114\% \\
w/o Caching & +9.6\% & +12.8\% \\
\hline
\end{tabular}
\caption{Ablation study: relative percentage change of PinDCO latency compared against PinDCO, when a specific optimization is removed.}
\Description{Ablation study: relative percentage change of PinDCO latency compared against PinDCO, when a specific optimization is removed. Removing any of the components causes significant latency regression.}
\label{tab:latency}
\end{table}

As discussed in Section \ref{Serving Efficiency}, we improve serving efficiency using caching and dynamic batching to accommodate high QPS. To quantify the contribution of each technique, Table \ref{tab:latency} reports P99 and P90 latencies when each optimization is removed.

Without dynamic batching, all candidates are concatenated into a single large scoring request, which is more susceptible to stragglers and loses parallelism for request/response processing. As shown in Table \ref{tab:latency}, latency increases by 87\% for P99 and 114\% for P90, highlighting the importance of dynamic batching.

Caching reduces feature fetches to the backend key-value store and enables low-latency access on cache hits. By optimizing the data access phase, it reduces latency by 9.6\% for P99 and 12.8\% for P90, though not directly accelerating model inference.

\subsubsection{Effect of Pixel-aware Adjustment Module}

\begin{table}[h]
\centering
\begin{tabular}{lcc}
\hline
PAM & DCO Ad CTR & DCO Ad Aspect Ratio \\
\hline
No-CR & - & - \\
PinDCO & 9.5\% & 5.22\% \\
w/o PAM & 9.8\% & 7.21\% \\
\hline
\end{tabular}
\caption{Relative percentage change of DCO ad CTR and aspect ratio when keeping or removing Pixel-aware Adjustment Module(PAM), compared against the no-CR baseline.}
\Description{Relative percentage change of DCO ad CTR and aspect ratio when keeping or removing Pixel-aware Adjustment Module(PAM), compared against the no-CR baseline. The PAM greatly reduced pixel footprint while maintaining most of the CTR gains.}
\label{tab:arp}
\end{table}

The Pixel-aware Adjustment Module(PAM) discourages excessively long creatives to reduce cannibalization of other content. Table \ref{tab:arp} shows that it effectively constrains creative aspect ratio while preserving engagement gains. To undilute the effect, we focus on DCO ads with multiple creative variants. Without the PAM, CTR improves by $9.8\%$ with a $7.21\%$ increase in aspect ratio. When PinDCO applies the penalty, we still obtain a comparable $9.5\%$ CTR gain relative to no penalty, but with a smaller increase in aspect ratio ($5.22\%$ vs $7.21\%$). Overall, the penalty encourages more effective use of available pixels.

\subsubsection{Personalization}
The creative selection system allocates creative variants to the most suitable audiences. The resulting delivery patterns should align with user preferences. Accordingly, we bucket users and requests into distinct groups and analyze delivery patterns across these buckets.

\begin{figure}[h]
    \centering
    \includegraphics[width=0.5\textwidth]{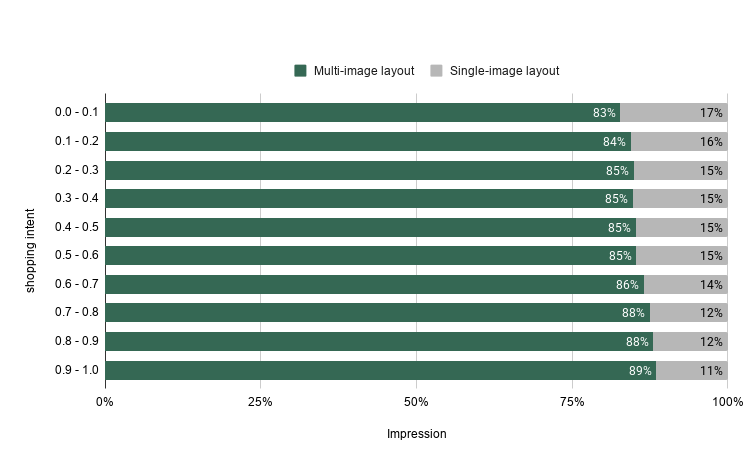}
    \caption{Impression breakdown of layout by user shopping intent.}
    \Description{Impression breakdown of layout by user shopping intent. multi-image layout is more common for beauty, fashion, and diy crafts, but less common for architecture or design.}
    \label{fig:shopping-intent-breakdown}
\end{figure}

\begin{figure}[h]
    \centering
    \includegraphics[width=0.5\textwidth]{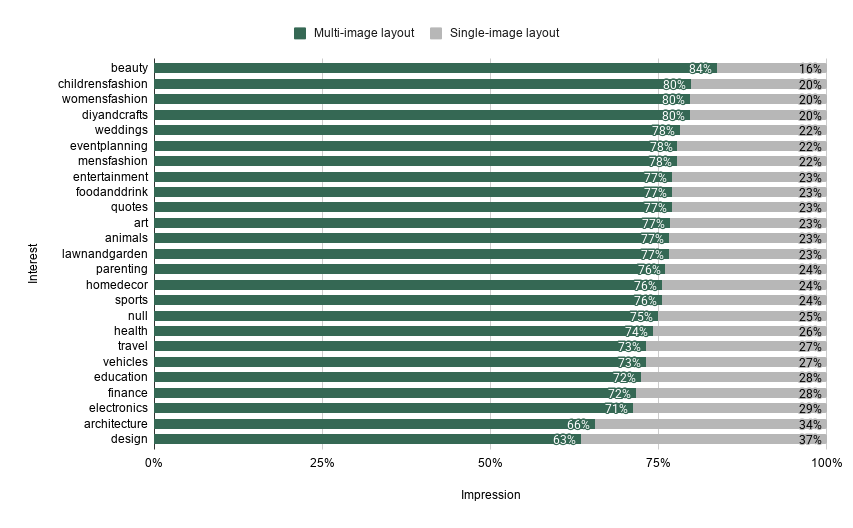}
    \caption{Impression breakdown of layout by user search query interest.}
    \Description{Impression breakdown of layout by user search query interest. Users with stronger shopping intent show a higher multi-image rate}
    \label{fig:interest-breakdown}
\end{figure}

As an illustrative example, in Figures \ref{fig:shopping-intent-breakdown} and \ref{fig:interest-breakdown}, we group impressions into buckets based on query interest(Pinterest internal category of topics) and user shopping intent, which is a signal derived from user behavior. We then compare the impression distribution between two layouts, multi-image and single image, within each bucket.

The multi-image layout places a hero image on top with smaller drawer images below; typically the hero shows the main product and the drawers show comparable items. In shopping-centric sessions, it can increase product exposure and ease comparison, improving engagement by better matching user preferences. In Figure \ref{fig:interest-breakdown}, multi-image impressions are more common for beauty, fashion, and diy crafts, which often feature product-focused creatives suited to this showcase layout. Consistently, users with stronger shopping intent show a higher multi-image rate in Figure \ref{fig:shopping-intent-breakdown}.

In contrast, the single-image layout provides a more focused and immersive visual experience that emphasizes taste, aesthetic, and inspiration. Accordingly, in Figure \ref{fig:interest-breakdown}, the single-image layout is rendered more frequently for architecture and design than for other interests. In Figure \ref{fig:shopping-intent-breakdown}, the single-image layout is shown more often to users with lower shopping intent to preserve pixels for other content.

\begin{figure}[h]
    \centering
    \includegraphics[width=0.4\textwidth]{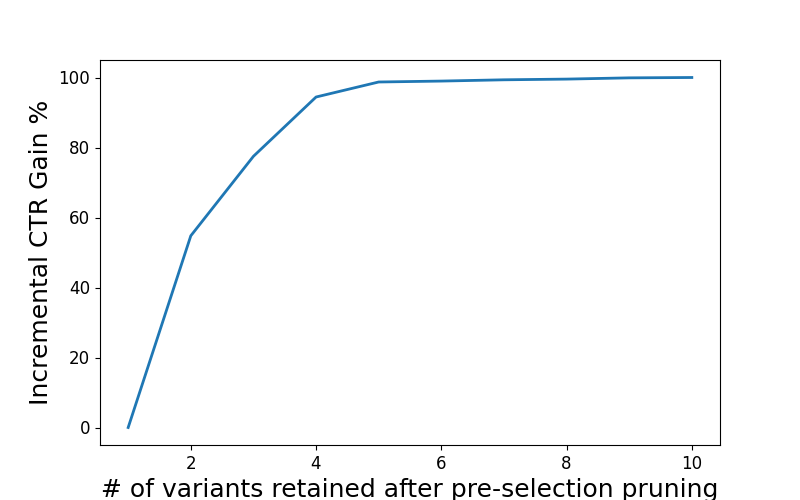}
    \caption{Realized CTR gain percentage by the number of variants retained after pre-selection pruning.}
    \Description{Realized CTR gain percentage by the number of variants retained after pre-selection pruning. We observe a clear monotonic improvement with reduced margin as more variants are retained.}
    \label{fig:fan-out}
\end{figure}

\subsubsection{Early Pruning}

Figure \ref{fig:fan-out} characterizes how the benefit of creative optimization varies with the number of variants retained after pre-selection pruning. We report the incremental CTR gain and normalize it with the maximum gain.

We observe a clear monotonic improvement with reduced margin as more variants are retained: The gain first rises rapidly when increasing the retained variant count, after which the curve begins to plateau. This pattern suggests that increasing the candidate set provides the CCFN model with more opportunity to identify a better-matched creative for the current user and context, but that the marginal return diminishes once enough diversity has already been preserved. The results therefore support the use of lightweight pre-selection to choose an operating point that captures most of the available performance gain while maintaining serving efficiency at scale. 

\section{Conclusion}
We present PinDCO, a Dynamic Creative Optimization system that efficiently matches a continually expanding repository of creative variants to appropriate audiences. The system comprises multiple components. The creative asset manager expands an ad into its creative variants, which are subsequently pruned by a pre-selection module to reduce inference cost. The Creative Component Fusion Network serves as the core scoring model, using creative-component-specific towers and hyperparameters to account for component complexity. We further adjust creative scores using an Pixel-aware Adjustment Module to balance slot-level engagement with overall whole-page impact, particularly under Pinterest's waterfall grid layout. Finally, we deploy a multi-armed bandit to explore and collect higher-quality training data. We evaluate PinDCO through extensive offline and online studies. With an online lift of $+3.09\%$ in platform ad CTR, we launched PinDCO on the Pinterest Ads Platform.

In future work, we aim to improve the expressiveness of the creative scoring model by advancing the model architecture, enriching creative signals, and incorporating user sequence modeling. Another important direction is to better unify creative asset generation and selection. A tighter integration could more directly leverage the user preference signals that PinDCO learns from real traffic, which may more efficiently guide the generator to produce more diverse and engaging creatives tailored to different audiences.

\begin{acks}
We would like to thank Meredith Martz, Enbo Zhou, Chun-Han Lu, Mukesh Bangalore Renuka, Fang He, Zihao Zhang for their contributions to the project.
\end{acks}

\bibliographystyle{ACM-Reference-Format}
\bibliography{main}

\end{document}